# X-Ray Microscopy of Spin Wave Focusing using a Fresnel Zone Plate

Joachim Gräfe[1*], Martin Decker[2], Kahraman Keskinbora[1], Matthias Noske[1], Przemysław Gawronski[3], Hermann Stoll[1], Christian H. Back[2], Eberhard J. Goering[1], Gisela Schütz[1]

[1]Max Planck Institute for Intelligent Systems, Stuttgart, Germany
[2]Department of Physics, University of Regensburg, Regensburg, Germany
[3]AGH University of Science and Technology, Krakow, Poland

*e-mail: graefe@is.mpg.de

*Magnonics, i.e. the artificial manipulation of spin waves, is a flourishing field of research with many potential uses in data processing within reach [1-4]. Apart from the technological applications the possibility to directly influence and observe these types of waves is of great interest for fundamental research. Guidance and steering of spin waves has been previously shown [5, 6] and lateral spin wave confinement has been achieved [7, 8]. However, true spin wave focusing with both lateral confinement and increase in amplitude has not been shown before. Here, we show for the first time spin wave focusing by realizing a Fresnel zone plate type lens. Using x-ray microscopy we are able to directly image the propagation of spin waves into the nanometer sized focal spot. Furthermore, we observe that the focal spot can be freely moved in a large area by small variations of the bias field. Thus, this type of lens provides a steerable intense nanometer sized spin wave source. Potentially, this could be used to selectively illuminate magnonic devices like nano oscillators [9] with a steerable spin wave beam.*

Confinement of spin wave beams by lateral restriction of the propagation medium [8] and self-collimation [10] has been already demonstrated. However, both processes occur at a loss of spin wave intensity. A real lens that is able to focus spin waves, *i.e.* yield and increase of intensity in a confined spot, has not been realized so far [11], especially so in an unstructured thin film. As an array of holes acts as a diffraction lattice for spin waves [12], a hole arrangement seems to be a promising base for the realization of a diffractive lens. A Fresnel zone plate is such a diffractive lens that is widely used in focusing of light [13]. For spin waves this can be realized by an alternating arrangement of material (transmissive) and holes (absorbing), as shown in Figure 1. This design was calculated using a Rayleigh Sommerfeld propagator [14] assuming the wavelength of Damon-Eshbach type spin waves that were previously observed. Thus, neglecting the anisotropic dispersion properties of spin waves, but resulting in a sufficient design approximation to operate as proof of concept of a spin wave lens.

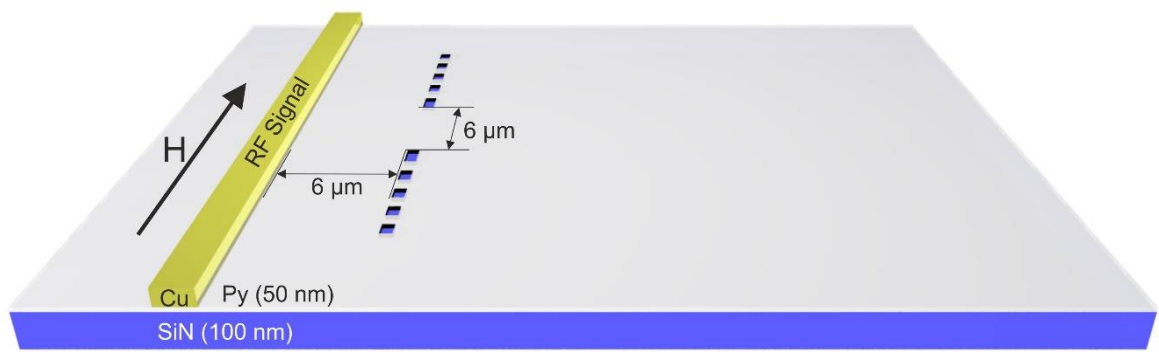

*Figure 1: Sketch of the magnonic zone plate samples for x-ray microscopy. 50 nm of Py are deposited on top of a x-ray transparent $Si_3N_4$ membrane. For RF excitation a 1.6 µm wide Cu stripline is deposited on top of the magnetic thin film as excitation source. The Fresnel zone plate is realized by an arrangement of holes at a distance of 6 µm from the excitation source. Additionally, an in-plane bias field is applied along the stripline.*

## Focusing Characteristics

Spin waves passing through such a Fresnel zone plate were imaged by time resolved scanning x-ray transmission microscopy with XMCD [15] contrast. This technique visualizes the spatial distribution of the time dependent magnetization perpendicular to the sample surface. The spin wave amplitude derived from these data is shown in Figure 2 (*cf.* Supplemental 1 and 2 for spin wave movies at 3.6 and 5.7 GHz respectively). While Figure 2 shows the spin wave amplitude at 3.6 GHz this lens geometry is also capable of focusing spin waves at 5.7 GHz (shown in Supplemental 3). Thus, indicating that this type of lens is capable of operating in a broad frequency range. Areas with high spin wave amplitude are shown in bright colors, and areas with low amplitude are shown in black.

Here, an amplitude increase of 23% at a focal distance of 6 µm behind the lens is found, thus, overcompensating the spin wave damping during propagation in permalloy. This occurs at a confinement of the spin wave beam to 840 nm FWHM. In the focal plane, the focal spot features a 10:1 raised amplitude within the central region behind the zone plate. Furthermore, the focal spot stands out with a 2:1 raised amplitude compared to the whole interference pattern. Although this zone plate design only features a limited number of zones, this clearly shows that focusing of spin waves is possible and the focal spot is significantly standing out of the spin wave landscape.

A similar behavior is found at a higher frequency of 5.7 GHz. At these frequencies the focal length is reduced to 4.5 µm (shown in Supplemental 3). Compared to lower frequencies the focal width is slightly increased to 1.1 µm FWHM in our measurements. Also in this case, there is a strong enhancement of the spin wave amplitude in the focal spot within the central region behind the zone plate and the focal spot stands out of the whole interference pattern behind the zone plate, resulting in the most intense spot in the focal plane.

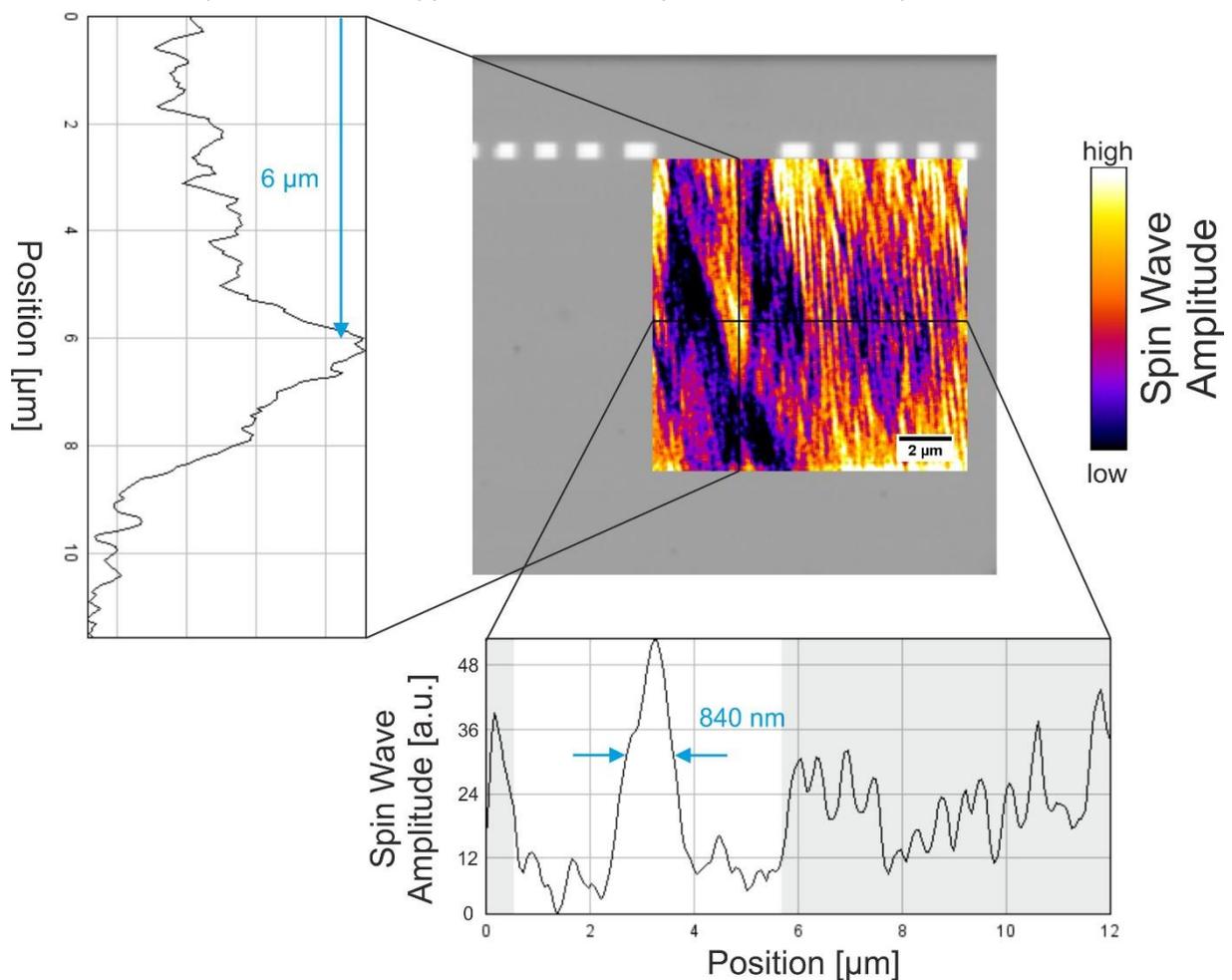

*Figure 2: Spin wave amplitude for transmission through a hole based Fresnel zone plate in a 50 nm Py thin film at 3.6 GHz with an applied in-plane bias field of -12 mT. Additionally, cuts through the amplitude along the propagation direction and through the focal plane are shown. In the focal spot at 6 µm behind the lens the spin waves are confined to 840 nm FWHM. Furthermore, the amplitude is increased by 23%, thus, overcompensating for damping during spin wave propagation.*

Furthermore, the spin wave damping during propagation was approximated by fitting an exponential decay to the overall spin wave amplitude along the propagation direction. When considering damping, an amplitude enhancement of 80% is reached in the focal spot in comparison to the source. This is equal to the amplitude enhancement that is calculated from a conventional Rayleigh-Sommerfeld propagation approach [14]. This suggests that all zones contribute to the spin wave amplitude in the focal region and here, indeed, diffractive focusing of spin waves is observed. However, the Rayleigh-Sommerfeld propagation fails to describe the focal length correctly, because the anisotropic dispersion [3] of spin waves in a uniform medium is neglected. Therefore, the system has to be modeled by micromagnetic simulations (*cf.* Supplemental 4).

Another peculiarity is the small focal width in comparison to the wavelength of several µm of the Damon-Eshbach type spin waves observed here. In principle the focal width is significantly smaller than the wavelength of the spin waves that are being focused. However, this is not real sub-wavelength focusing due to the before mentioned anisotropic dispersion relation [3]. The wavelength in the direction perpendicular to the propagation (Backward Volume geometry) is significantly shorter at the same frequency, thus, the focal width is not smaller than the wavelength along the focal line.

## Focus Steering

Unlike zone plates for photons, the focal position of the zone plate for spin waves is not fixed for a given frequency. In fact, the focal position can be moved within a 40 µm² area behind the zone plate by changing the applied static field as shown in Figure 3. At low (-6 mT, Figure 3a) and high (-40 mT, Figure 3h) field amplitudes no focusing occurs, while the focal position is shifted from the outer edge of the central area to its center (Figure 3b-g) at intermediate field amplitudes. The same behavior can be observed for opposing field sign with inverted symmetry.

This can be understood by a field dependent mixing of Damon-Eshbach ($\vec{k} \perp \vec{M}$) and Backward Volume ($\vec{k} \parallel \vec{M}$) type spin waves at a given frequency [4]. While the former are propagating perpendicular to the applied field, and thus, to the zone plate, and have long wavelength; the latter are oriented parallel to the applied field and feature shorter wavelengths.

For spin waves with large Damon-Eshbach type contribution (Figure 3d-g) the focal length is long and the focal spot lies roughly in the center behind the zone plate. On the other hand, for spin waves with large Backward Volume type contribution (Figure 3c+d) the focal length is shorter, due to the shorter wave length, and the focal spot lies further outwards, because of the propagation component parallel to the external field. Furthermore, the k-vector of the Backward Volume type spin waves points in opposing directions when the sign of the applied field is flipped, and thus, the focal shift occurs in the other direction from the center.

This shortening of the wavelength and the change in propagation direction is easily observed in the phase contrast of Figure 3. This phase information at high resolutions is uniquely available from X-ray microscopy measurements that allow the experimental explanation of the spin wave zone plate properties.

In this context, the focusing cut off at low and high field amplitudes can be understood as well. At high fields the large Damon-Eshbach type contribution results in very long spin wave wavelength and the spin waves are damped out before constructive interference results in the formation of a focal spot. At low fields, on the other hand, the large Backward Volume type contribution results in a very shallow incidence angle of the spin waves on the zone plate. The cut off occurs when the spin waves cannot pass through the outermost zones and constructive interference is, therefore, inhibited. This additionally confirms that all zones contribute to the spin wave amplitude within the focal spot.

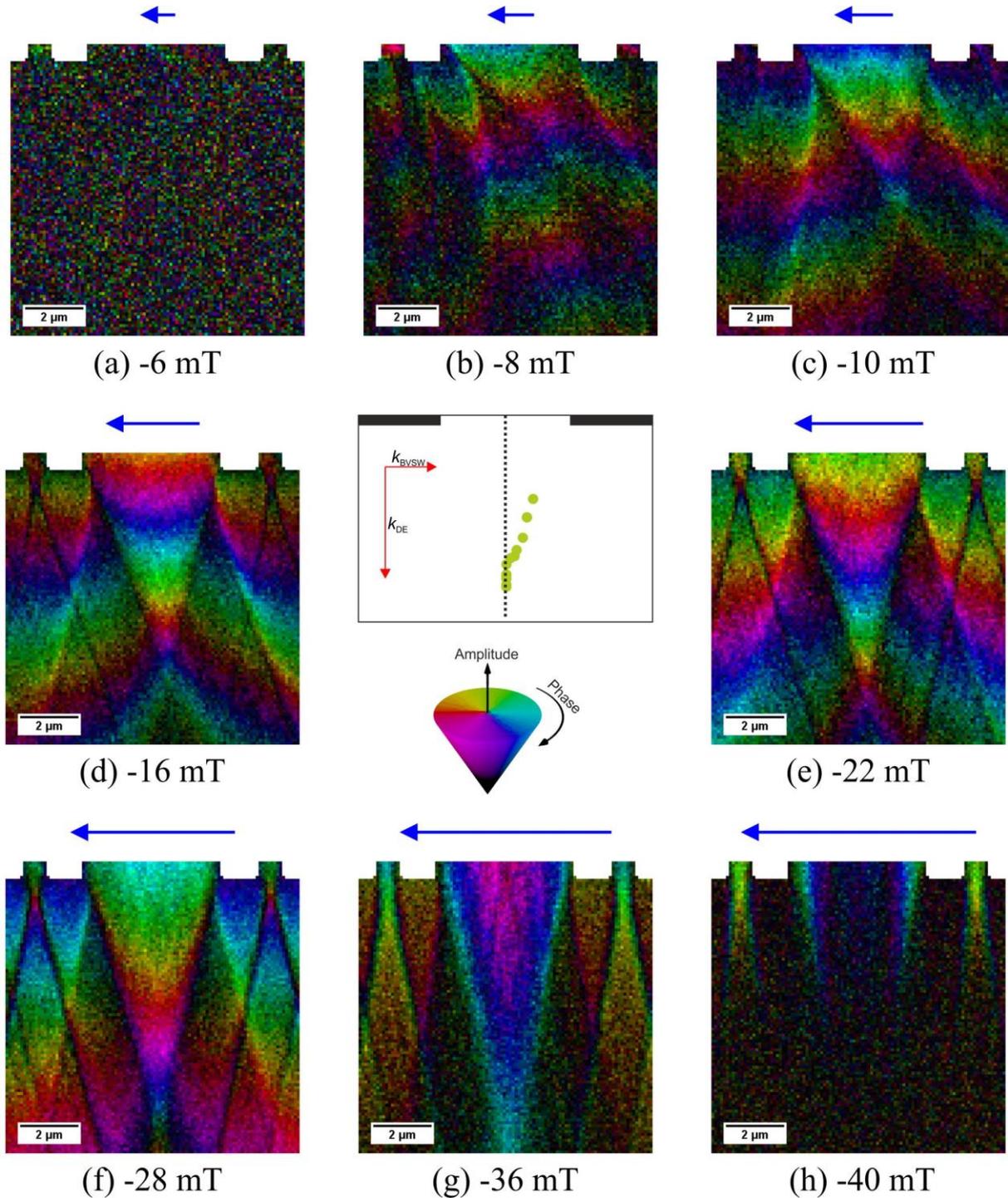

Figure 3: Spin wave amplitude and phase for transmission through a hole based Fresnel zone plate in a 50 nm Py thin film at 5.7 GHz with applied in-plane bias fields. At field magnitudes below 6 mT (a) or above 40 mT (h) no spin wave focusing occurs. At intermediate fields the focal spot moves inwards and downwards with increasing Damon-Eshbach type spin wave contribution, i.e. the wave length increases and the emission angle turns perpendicular to the external field. The central inset shows the focal positions (maximum lateral confinement) as green dots for fields ranging from -8 mT to -36 mT.

## Conclusion

In summary, we have shown for the first time that a true lens for focusing spin waves can be realized by a Fresnel zone plate design that is capable of broad band operation. Here, our diffractive focusing approach results in a focal spot with 840 nm FWHM at a 23% increase of spin wave amplitude, overcompensating damping during propagation. Furthermore, the focal spot can be moved within a 40 µm² area behind the zone plate by varying the external magnetic field.

The behavior of focus formation and shifting with applied field are explained by field dependent mixing of Damon-Eshbach and Backward Volume type contributions to the observed spin waves. Depending on the wavelength and propagation direction, different focal positions are achieved. Furthermore, this explains the focusing cut off at low (shallow incidence angle) and high (long wavelength) field amplitudes.

The possibility of focusing spin waves and steering the focal position opens up a large number in magnonics research. This first realization of a lens for spin wave provides a highly localized and intense spot of spin waves within a uniform magnetic film. Thus, allowing any devices to be placed into the focal position without limitations imposed by the focusing system. Furthermore, the nanometer sized spin wave illumination can easily be switched off a device or switched between several devices placed into the movability area of the focal spot. For example a magnonic logic device based on multiple spin wave synchronized nano oscillators [9] is easily conceived, where specific output channels could be addressed by changing the spin wave position. Effectively, creating a spin wave distributor system.

## Methods

50 nm thin permalloy (Py, $Ni_{80}Fe_{20}$) films were deposited on $Si_3N_4$(100 nm)/Si(100) substrates by evaporation at a base pressure better than $1 \cdot 10^{-7}$ mbar and covered by 5 nm thin Al. The holes making up the zone plates where structured by e-beam lithography and lift-off. Subsequently, 10 nm of $Al_2O_3$ where deposited by atomic layer deposition for isolation of a 1.6 µm wide microstrip antenna [Cr(10 nm)/Cu(150 nm)/Al(5 nm)] that was deposited on top.

Time resolved STXM measurements were conducted at the MPI IS operated MAXYMUS end station at the UE46-PGM2 beam line at the BESSY II synchrotron radiation facility. The samples were illuminated under perpendicular incidence by circularly polarized light in an applied in-plane filed of up to 240 mT that was generated by a set of four rotatable permanent magnets [16]. The photon energy was set to the absorption maximum of the Fe $L_3$ edge to get optimal XMCD contrast for imaging. A lock-in like detection scheme allows sample excitation at arbitrary frequencies at a time resolution of 50 ps using all photons emitted by the synchrotron.

## Acknowledgment

The authors would like to thank Michael Bechtel for support during beam times. Johannes Stigloher is gratefully acknowledged for fruitful discussions. Helmholtz Zentrum Berlin is acknowledged for allocating beam time at the BESSY II synchrotron radiation facility. Financial support by the Baden-Württemberg Stiftung in the framework of the Kompetenznetz Funktionelle Nanostrukturen is gratefully acknowledged. The work of P. G. was partially supported by the PL-Grid infrastructure.

## Supplemental 1

MPI_150529029.mp4

*Spin wave movie for transmission through a hole based Fresnel zone plate in a 50 nm Py thin film at 3.6 GHz with an applied in-plane bias field of -12 mT.*

## Supplemental 2

MPI_150929027.mp4

*Spin wave movie for transmission through a hole based Fresnel zone plate in a 50 nm Py thin film at 5.7 GHz with an applied in-plane bias field of -18 mT.*

## Supplemental 3

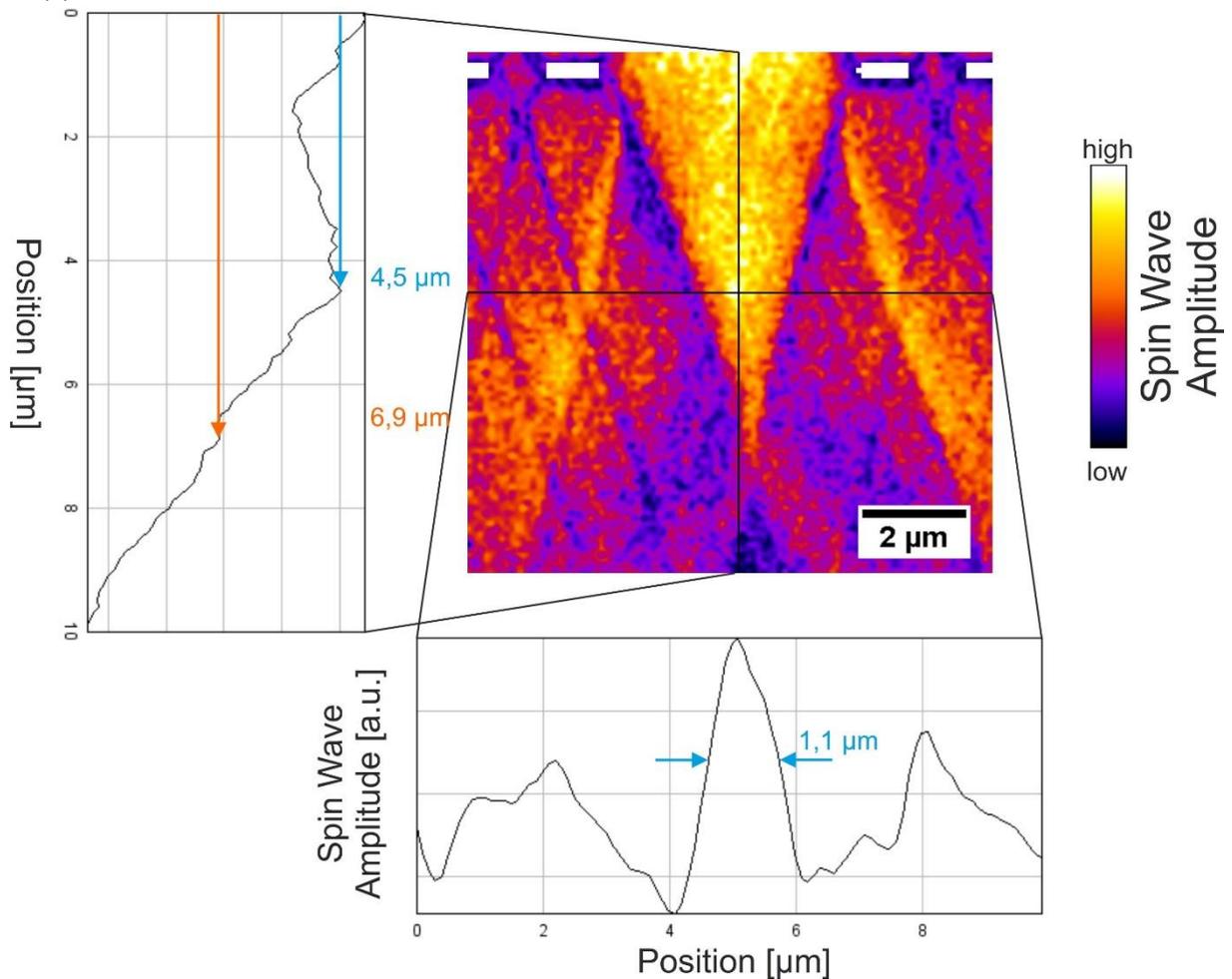

*Spin wave amplitude for transmission through a hole based Fresnel zone plate in a 50 nm Py thin film at 5.7 GHz with an applied in-plane bias field of -18 mT. Additionally, cuts through the amplitude along the propagation direction and through the focal spot are shown. In the focal spot (maximum intensity) at 4.5 µm behind the lens the spin waves are confined to 1.1 µm FWHM. The maximum lateral confinement of spin waves occurs at 6.9 µm behind the lens.*

Supplemental 4

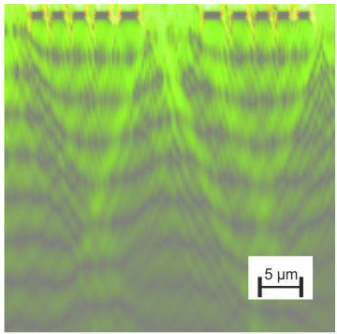 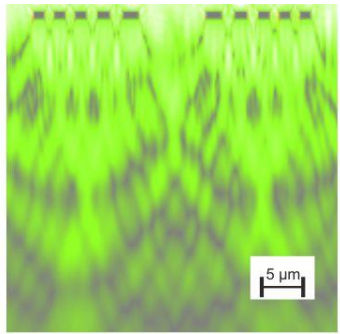 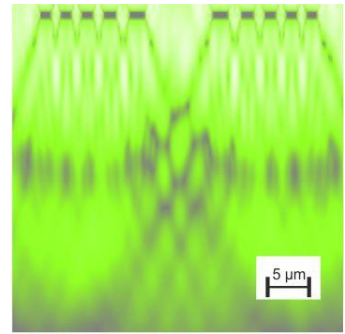

(a) +3 mT    (b) +12 mT    (c) +15 mT

*Micromagnetic simulation of spin wave amplitude for transmission through a hole based Fresnel zone plate in a 50 nm Py thin film at 3.6 GHz with an applied in-plane bias field of (a) +3 mT, (b) +12 mT, and (c) +15 mT.*